# Influence of the formation of clusters on the effective elastic properties of platelet reinforced polymers

Sergejs Tarasovs*, Andrey Aniskevich

*Institute for Mechanics of Materials, University of Latvia, Jelgavas st. 3, Riga, LV-1004, Latvia*

* – Corresponding author: tarasov@pmi.lv

**Abstract**

The effect of the clusterization on the effective properties of a composite material reinforced by MXene or graphene platelets is studied using the finite element method with periodic representative volume element (RVE). A hybrid 2D/3D finite element mesh is used to reduce the computational complexity of the numerical model. Several realizations of an RVE were generated with increasing volume fractions of inclusions, resulting in progressive clusterization of the platelets. Numerically obtained effective properties of the composite are compared with analytical predictions by the Mori-Tanaka method and Halpin-Tsai equations, and the limits of the applicability of the analytical models are established. A two-step homogenization scheme is proposed to increase the accuracy and stability of the effective properties of an RVE with a relatively small number of inclusions. Simple scaling relations are proposed to generalize numerical results to platelets with other aspect ratios.

## 1 Introduction

Polymer composites reinforced by stiff thin platelets have been studied extensively in recent years as a new generation of materials. Unique properties of platelets, such as the high aspect ratio, stiffness, and strength, offer a significant improvement of mechanical properties of the composite material even at a very low volume fraction of reinforcing inclusions. However, the effective properties of a composite strongly depend on the technological process of the preparation of reinforcing nanoparticles and the quality of dispersion. Densely packed aggregates of particles are often observed experimentally, which has to be considered in analytical or numerical models when studying the effective properties of the composite material.

For polymer/clay composites three distinct shapes of particles are often considered (Giannelis, 1996; Lan et al., 1995; Messersmith and Giannelis, 1994):

    a)   conventional particle (relatively thick layered silicate particle);

---

b) intercalated particles, with layers separated by polymeric molecules, having a shape of "swollen" microaggregates;
c) exfoliated particles, with layers, fully separated by polymer matrix, however sometimes still in the shape of a layered stack, but with much larger inter-layer distance than in case of intercalated particles.

Similar morphology can be observed in MXenes with a MAX phase as a conventional particle, and intercalated and exfoliated particles after etching and delamination procedures (Halim et al., 2016; Monastyreckis et al., 2020; Naguib et al., 2014).

Exfoliated particles may form aggregates of different shapes:

a) "House of Cards", where neighboring particles are in a "T"-shape configuration;
b) "Stacked Plates", where particles are in a sandwich configuration;
c) narrow strips of "Overlapping Coins".

Different combinations of such shapes and aggregates are also possible and have been observed experimentally. As the volume fraction of particles increases, stiff particles may form clusters of few or many nearly parallel platelets, as schematically is shown in Figure 1. The detailed understanding of how different morphology of reinforcing particles affects the properties of composite material is still far from complete, and many experimental and numerical investigations were performed in recent years.

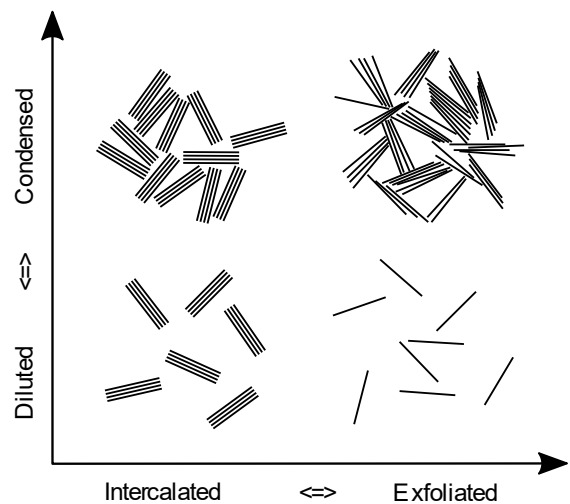

Figure 1 Examples of the arrangement of intercalated and exfoliated platelets in diluted and condensed states

Wang et al. (2005) studied the elastic modulus, tensile strength, and fracture toughness of an epoxy/clay composite exfoliated in a "slurry-compounding" process. The exfoliated clay particles in their experiments had a form of tactoids that contain only a few clay layers separated by polymer matrix, and these tactoids were dispersed uniformly in the matrix, creating a "House of Cards" network.

Zilg et al. (1999) have investigated the influence of the nanocomposite morphology and the formation of a structure on the toughness and stiffness of the material. Many different compositions were investigated with both intercalated and fully exfoliated layered clay particles.

Molecular dynamics (MD) simulations of a consolidation process of charged clay particles in a colloid suspension and a formation of clusters were performed in Delhorme et al. (2012) and Ebrahimi et al. (2016). In the work of Ebrahimi et al. (2016), particles of different aspect ratios and sizes were studied, and an average size of clusters was determined in MD simulations for monodisperse and polydisperse compositions of platelets as a function of increasing pressure. In the work of Coelho et al. (1997) the sequential deposition of particles in the "gravitational" field was used to study a random packing of particles of an arbitrary shape. The Particle Flow Code was used in Bock et al. (2006) to study the reorientation of platelets during a uniaxial compaction process.

An extensive parametric study of intercalated, exfoliated, and partially exfoliated (hybrid) polymer/clay nanocomposites with a volume fraction of up to 2% was performed in Vo et al. (2018). The influence of agglomeration of graphene sheets on the effective elastic properties was studied by Ji et al. (2010) employing a two-step Mori-Tanaka homogenization approach. An effect of various parameters, like the aspect ratio of particles or interphase stiffness and thickness, on the effective stiffness of composite material and the micro-strain level in the matrix phase was studied by Hussein and Kim (2018). A direct relation between the orientation tensor anisotropy and the anisotropy of the effective stiffness matrix was observed.

Peng et al. (2012) developed a 2D finite element model for particles with interphase to study the effect of a distribution of particles on overall elastic properties. In particular, the effect of clustering was studied for an "intercalated" arrangement of elliptical particles.

Hbaieb et al. (2007) studied the effect of the clustering on the effective stiffness using a 2D finite element model and found that 2D models do not accurately predict elastic properties of composites reinforced by platelets. The dense clusters forming at high volume fractions of inclusions affect the reinforcing ability of the platelets, as a result, the predictions of the Mori-Tanaka method overestimate the effective elastic properties of the composite.

An extensive numerical and analytical (by Mori-Tanaka method) study was performed by Figiel and Buckley (2009) for a layered-silicate/polymer nanocomposite using an effective particle concept, where intercalated stacks of parallel platelets are first replaced by a larger homogenized isotropic or anisotropic "effective particle". The effect of the elastic anisotropy of the individual effective particles and the degree of orientation of such particles in the composite material was studied.

The aim of this work is to study the influence of the formation of dense clusters on the effective elastic properties of a composite reinforced by thin stiff platelets using a 3D representative volume element (RVE) with a large number of fully exfoliated particles. The optimal size of the RVE was evaluated using numerical models with 159, 310, and 464 particles. The numerical results are compared with analytical predictions by the Mori-Tanaka method and Halpin-Tsai equations, and limits of the applicability of analytical models are established.

The interphase effect was not considered in this study. As was shown (Mortazavi et al., 2013a), interphase plays a small role for the disk-shaped reinforcement with a high aspect ratio, and the reinforcing role of the interphase decreases as the aspect ratio increases. For typical geometries and properties of graphene and MXene particles, the reinforcing effect of the interphase in the order of a few percent was obtained by numerical simulations (Kilikevičius et al., 2020).

# 2 Numerical and analytical models

## 2.1 Generation of an RVE

The arrangement of particles inside an RVE was generated by the in-house developed computer code. Each particle was treated as a circular disk with a finite thickness (to allow both 2D and 3D finite element models of the filler) embedded in a unit cube. A modified Random Sequential Adsorption (RSA) algorithm was used to generate positions and orientations of particles. A two-step process was implemented:

1) In the first step the classical RSA (Feder, 1980; Widom, 1966) algorithm was used to generate an initial geometry with a reasonable low volume fraction of a filler, resulting in a dilute system of weakly interacting particles;
2) During the second step, an iterative process was employed to increase the size of the particles, until the desired volume fraction of the inclusions is achieved. At each iteration the following procedures were performed:
    a. All particles were scaled by a small fraction retaining their positions and orientations.
    b. An overlap check was performed for each pair of particles, and if particles were overlapping, one particle was shifted and rotated by a small amount to find a new non-overlapping position.
    c. If, after several attempts, all particles were successfully placed into new non-overlapping positions, the new configuration was accepted, and the next iteration was started; otherwise, the arrangement of particles was rolled back to the previous iteration and a "shake" step was performed.
    d. During the "shake" step, each particle consequently was shifted by a small amount to a new position and the overlapping criterion was checked. In case of success, the new position was accepted, otherwise, the particle was moved back, and a new attempt was performed. In case of several unsuccessful attempts, the old position was retained, and the cycle was continued for the next particle. The next iteration was started after the "shake" step.

During all transformations, the condition of a particle crossing the boundary of an RVE was checked and mirror particles were generated on the opposite sides of the RVE for the particles located on the boundary.

Finally, when the requested volume fraction was achieved, the whole RVE was translated several times by a random vector (keeping the periodicity of the geometry), and the minimum distance $\delta$ from each particle to the walls, edges, and vertices of the RVE was computed. The configuration with the maximum value $\delta$ was used for analysis to avoid excessive distortion of finite elements (Vila-Chã et al., 2021).

The procedure described above allowed generating an RVE with high volume fractions of thin platelets, when particles form dense clusters.

The geometry of the RVE was saved as a GEO script to be used by Gmsh (Geuzaine and Remacle, 2009) software. The geometrical engine of Gmsh was used to clip the particles by the RVE boundaries and then a periodic finite element mesh of the RVE was generated. The finite element mesh was exported to Ansys, where periodical boundary conditions (Li, 2001; Li and Wongsto, 2004) were enforced for each opposite pair of nodes by the in-house developed APDL script.

## 2.2 Estimation of the effective properties

Once the periodical boundary conditions are applied, the RVE can be loaded by an arbitrary combination of average strains or stresses, or the sides of the RVE can be left unloaded in some directions. This allows calculating a compliance matrix of the composite material directly by performing six independent solutions: tension and shear in three orthogonal directions. In each set of loads, only one stress component was applied at a time, and the deformation of the RVE was not restrained, allowing the RVE to freely contract due to a Poisson effect or to deform due to tension-shear coupling terms in the compliance matrix in case of an anisotropic material. The average strains of the RVE obtained in each of six analyses allowed the calculation of six components of the compliance matrix $s_{ij}$ (in Voigt notation) as:

$$s_{ij} = \frac{\varepsilon_i}{\sigma_j}; \quad i,j = 1,2,\dots 6 \tag{1}$$

where $\sigma_j$ is the applied average stress component and $\varepsilon_i$ are the resultant average global strain components. Alternatively, the compliance matrix can be written in the form:

$$s_{ij} = \begin{bmatrix} 1/E_x & -\nu_{yx}/E_y & -\nu_{zx}/E_z & \times & \times & \times \\ -\nu_{xy}/E_x & 1/E_y & -\nu_{zy}/E_z & \times & \times & \times \\ -\nu_{xz}/E_x & -\nu_{yz}/E_y & 1/E_z & \times & \times & \times \\ \times & \times & \times & 1/G_{yz} & \times & \times \\ \times & \times & \times & \times & 1/G_{xz} & \times \\ \times & \times & \times & \times & \times & 1/G_{xy} \end{bmatrix} \tag{2}$$

where $E_k$, $G_{kl}$ and $\nu_{kl}$ are the effective elastic constants, Young's modulus, shear modulus, and Poisson's ratio, respectively, measured in the directions of the local coordinate system of the RVE, and "×" are generally non-zero terms, approaching zero for a macroscopically isotropic RVE. By comparing the numerically calculated compliance matrix from Eq. (1) with its representation by Eq. (2), the effective elastic constants of a generally anisotropic material can be easily estimated.

## 2.3 Hybrid 2D/3D finite element mesh

Large-scale simulations of graphene, MXene or other thin planar inclusions embedded into polymer matrix require significant computational resources. The common practice of using 3D solid elements to model platelets leads to a very fine mesh and a large number of elements in case of inclusions with a high aspect ratio, limiting simulations of complex models with many particles. Simplified numerical models can be used to reduce required computer power, keeping the accuracy of the model at a satisfactory level. Spanos et al. (2015) used a hybrid finite element model, where beam elements were used to model graphene sheet, solid elements to model surrounding matrix, and the spring elements for an interface between the graphene and matrix.

In the current work, a hybrid 2D/3D mesh was used to model thin platelets embedded in an isotropic matrix. Graphene is an essentially 2D material and there is no widely accepted definition of its out-of-plane elastic properties. For other 2D materials, like MXenes, the out-of-plane properties can be determined, however, usually, only in-plane stiffness parameters are reported in the literature. The hybrid 2D/3D mesh was used to overcome this difficulty, where thin platelets were modeled as 2D finite elements embedded in the bulk of the matrix. The representation of the graphene sheet as a membrane or shell element is a natural choice, for other 2D materials with a finite thickness this approach should work if the aspect ratio of particles is sufficiently large, and the bending of the platelets could be neglected. Six-node triangular shell elements were used in the current work to model the platelets (Figure 2a), sharing nodes and faces with 10-node tetrahedral solid elements, representing the matrix (Figure 2b).

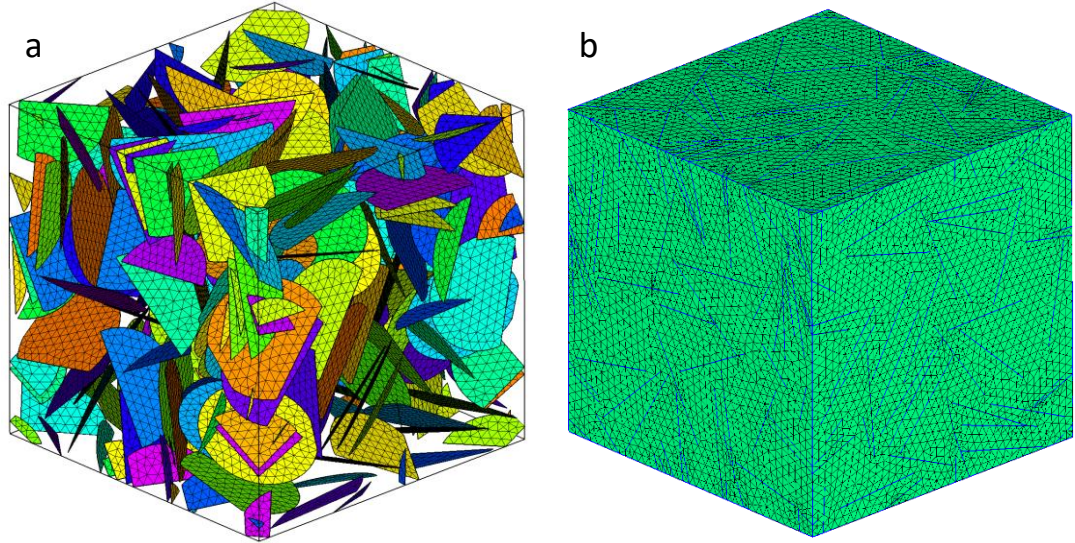

Figure 2 2D mesh of the reinforcing platelets (a) and the hybrid mesh of an RVE with embedded 2D inclusions (b)

## 2.4 Analytical models

Obtained numerical results were compared with predictions by two analytical models. Two commonly used methods for an estimation of the effective properties of composites reinforced by platelets, the Mori-Tanaka method (Benveniste, 1987; Mori and Tanaka, 1973) and the Halpin-Tsai equations (Affdl and Kardos, 1976), were used in this work. The Mori-Tanaka method, based on the Eshelby tensor for an ellipsoidal inclusion, is one of the most popular analytical methods for the calculation of the effective elastic properties of a composite reinforced by particles of an ellipsoidal shape, that can be reduced to a straight fiber or a planar disk for small or large values of an aspect ratio.

Another popular method for the estimation of the effective properties of composite material with a unidirectional distribution of inclusions is the Halpin-Tsai equations (3, 4):

$$\eta_L = \frac{E_f/E_m - 1}{E_f/E_m + \xi}; \quad \eta_T = \frac{E_f/E_m - 1}{E_f/E_m + 2} \tag{3}$$

$$\frac{E_L}{E_m} = \frac{1 + \xi \eta_L V_f}{1 - \eta_L V_f}; \quad \frac{E_T}{E_m} = \frac{1 + 2V_f}{1 - \eta_T V_f} \tag{4}$$

where $V_f$ is the volume fraction of the reinforcement, $E_m$ and $E_f$ are Young's moduli of the matrix and filler, respectively, and $\xi$ is the geometrical factor. The Halpin-Tsai equations are empirical but were successfully applied to many different inclusions by a proper choice of the geometrical factor $\xi$. For a material with a random three-dimensional dispersion of platelets, the effective stiffness can be calculated as (Van Es, 2001):

$$E_C = 0.49E_L + 0.51E_T \tag{5}$$

As it was shown (Van Es, 2001), with shape factor $\xi$ defined as

$$\xi = \frac{2D}{3h} \tag{6}$$

where $D$ is the diameter of the inclusion, $h$ is the thickness of the inclusion, the results of Halpin-Tsai equations are very close to the predictions by the Mori-Tanaka method for a wide range of parameters.

## 3 Results and discussions

### 3.1 Convergence study

To validate the accuracy of the hybrid 2D/3D numerical model, an example problem with stiff circular inclusions randomly dispersed in a soft matrix was considered. Figure 3 shows the comparison of the effective elastic properties of an RVE obtained in the FE models with the hybrid 2D/3D mesh (open circles), full 3D mesh (black diamonds), analytical predictions using the Mori-Tanaka method (solid line), and Halpin-Tsai equations (dashed line). In these simulations the diameter of particles $D$ (0.25 of the RVE size), inclusion-to-matrix stiffness ratio ($E_f/E_m = 10^6$, roughly corresponding to a graphene/silicone composite material), and the total number of particles ($N = 159$) were kept constant. The aspect ratio of particles was changed by reducing the thickness $h$, and, respectively, decreasing the volume fraction of the filler from 0.2 to $2 \cdot 10^{-8}$ as the aspect ratio $D/h$ was increased from 10 to $10^8$. The Poisson's ratio of the matrix and filler, $\nu_m$ and $\nu_f$ respectively, was equal to 0.3 for all simulations. Each numerical point in Figure 3 is an average of six random realizations of an RVE. The results show that the Halpin-Tsai equations with the shape factor $\xi$ defined in Eq. (6) practically coincide with the numerical solution obtained with the hybrid 2D/3D finite element mesh for a wide range of aspect ratios, except relatively thick platelets, where the three-dimensional shape of the inclusions cannot be neglected. For the particles with an aspect ratio less than or equal to 250, full 3D finite element models were generated, and the results are shown in Figure 3 as black diamonds, showing good agreement with both analytical models. The results of finite element simulations show that the hybrid solution converges to the full 3D solution for particles with aspect ratios greater than or equal to 250. Predictions by the Halpin-Tsai equations give satisfactory results even for quite thick platelets with an aspect ratio equal to 10.

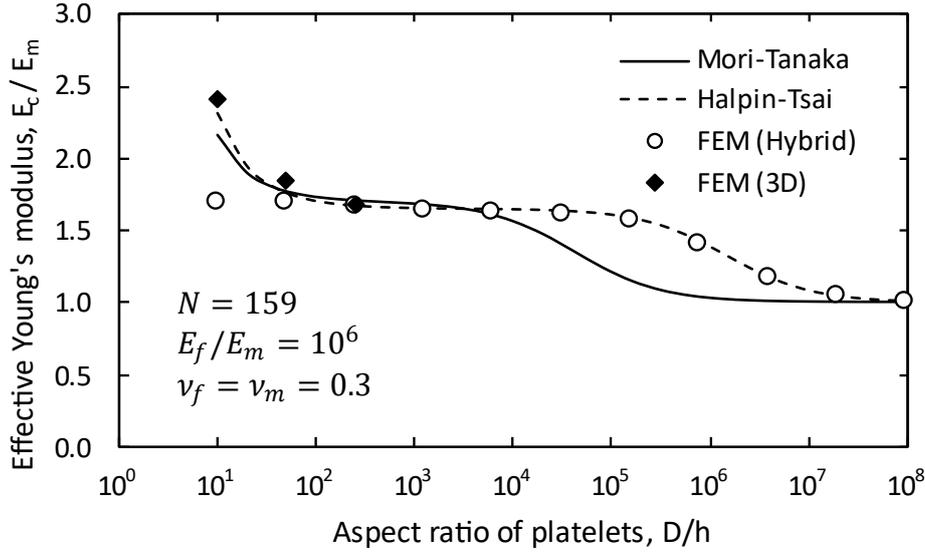

Figure 3 Convergence of the hybrid 2D/3D solution to a full 3D solution for various aspect ratios of platelets and comparison with the analytical solutions

## 3.2 The minimum size of an RVE

Previous studies have shown that an RVE with a relatively small number of randomly dispersed particles is needed to obtain an accurate estimation of the effective properties in the case of spherical inclusions. Segurado and Llorca (2002) used the RVE with 30 spherical particles and Gusev (1997) reported the accuracy of 3% and 7% for the stiffness matrix terms $c_{11}$ and $c_{12}$, respectively, with the RVE containing only 8 spheres. Disk shaped particles have lower symmetry and not only positions, but also orientations of particles should be considered, nevertheless Hbaieb et al. (2007) reported that numerical results are insensitive to the RVE size provided that more than 30 particles are used. Mortazavi et al. (2013b) used a 3D finite element model to estimate effective elastic and thermal properties of a platelet-reinforced polymer and concluded that 70 inclusions are sufficient to obtain accurate results. Vo et al. (2018) found that the effective stiffness is nearly independent of an RVE size when more than 50 particles were modeled. However, these estimations were obtained for dilute systems of particles. As the volume fraction of a filler increases, the inclusions form clusters of closely packed nearly parallel platelets. Such localized clusters may be considered as a larger "effective particle" with generally anisotropic elastic properties. Therefore, at least several randomly oriented clusters within an RVE are needed to achieve a macroscopically isotropic behavior of the modeled composite material and a much larger number of platelets should be used.

To study the formation of clusters in the compaction process and to estimate the influence of an RVE size on the effective elastic properties of the composite, three generations of an RVE with 159, 310, and 464 platelets were analyzed. An RVE with volume fractions of the filler ranging from 0.1% up to 3.0% (3.5% for the RVE with 464 particles) and the aspect ratio of platelets equal to 1000 were generated. Figures 4–6 show the examples of an arrangement of platelets within an RVE with 159, 310, and 464 particles, respectively, for different volume fractions of inclusions.

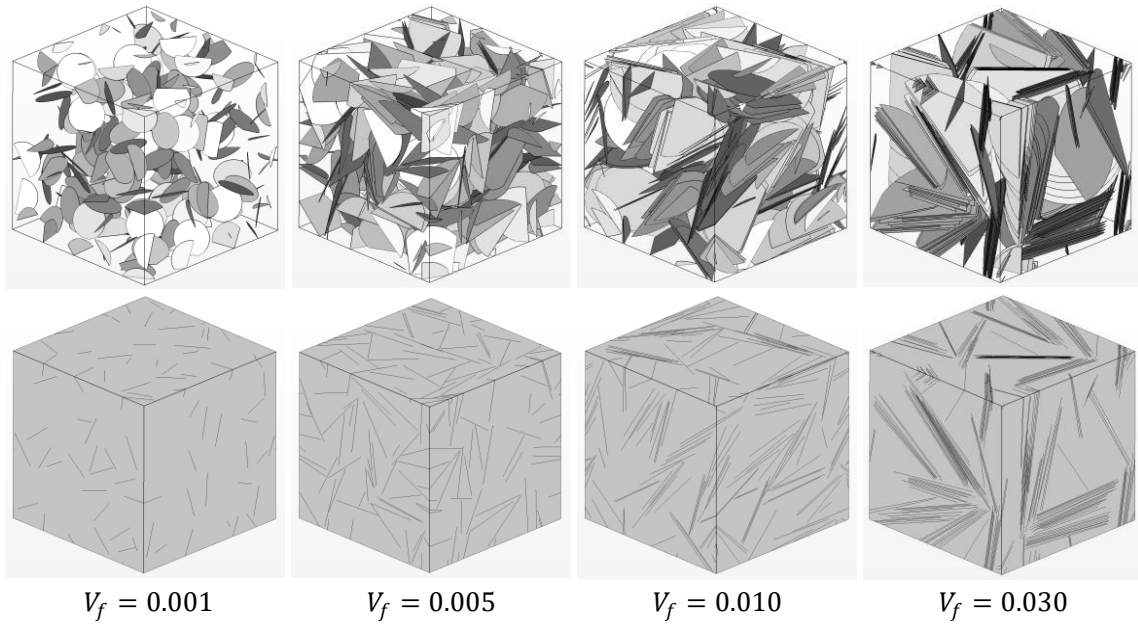

Figure 4 The arrangement of platelets in the RVE with 159 particles for different volume fractions (upper row) and the sides of the RVE with crossing particles (bottom row). The aspect ratio of particles (D/h) equals 1000

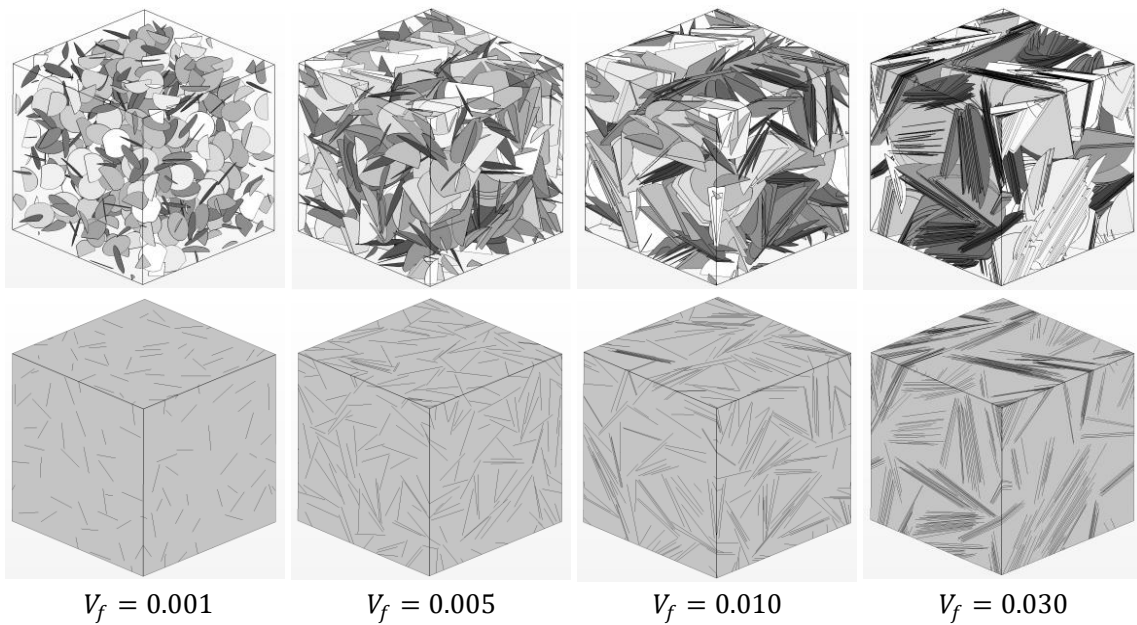

Figure 5 The arrangement of platelets in the RVE with 310 particles for different volume fractions (upper row) and the sides of the RVE with crossing particles (bottom row). The aspect ratio of particles (D/h) equals 1000

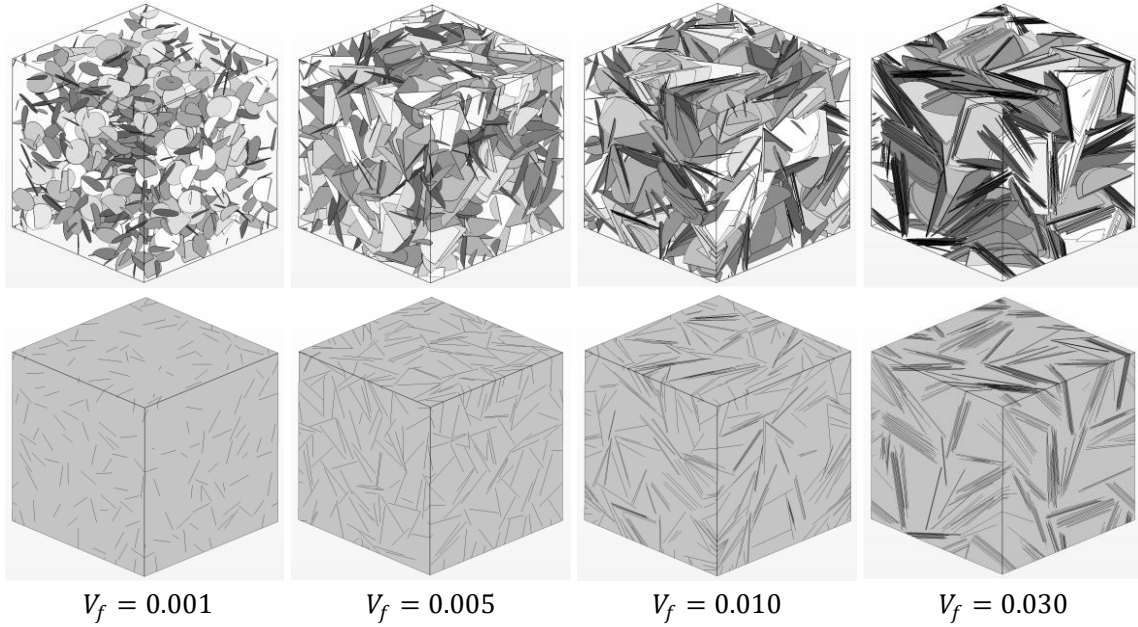

| $V_f = 0.001$ | $V_f = 0.005$ | $V_f = 0.010$ | $V_f = 0.030$ |

Figure 6 The arrangement of platelets in the RVE with 464 particles for different volume fractions (upper row) and the sides of the RVE with crossing particles (bottom row). The aspect ratio of particles (D/h) equals 1000

A degree of randomness in orientations of platelets can be estimated using a so-called average orientation tensor (Advani and Tucker III, 1987), defined as:

$$a_{ij} = \frac{1}{N}\sum_{k=1}^{N} n_i^k n_j^k, \qquad (7)$$

where $N$ is the total number of particles and $n_i^k$ are the components of a normal vector of $k$th particle. For an absolutely random distribution of a large number of platelets, the average orientation tensor is a diagonal matrix with the elements on the diagonal equal to $1/3$. For an RVE with a finite number of particles a degree of anisotropy, $A^G$, can be defined as a ratio of maximal and minimal diagonal elements of the tensor $a_{ij}$:

$$A^G = \frac{\max(a_{11}, a_{22}, a_{33})}{\min(a_{11}, a_{22}, a_{33})}. \qquad (8)$$

Figure 7 shows a box plot for the anisotropy of the orientation tensor of an RVE with 159, 310, and 464 particles and different volume fractions of inclusions. As could be expected, the higher is the number of particles, the more isotropic dispersion of platelets can be obtained. The degree of anisotropy grows very quickly for the RVE with 159 particles and high volume fractions of the filler, where most platelets are grouped into a small number of clusters.

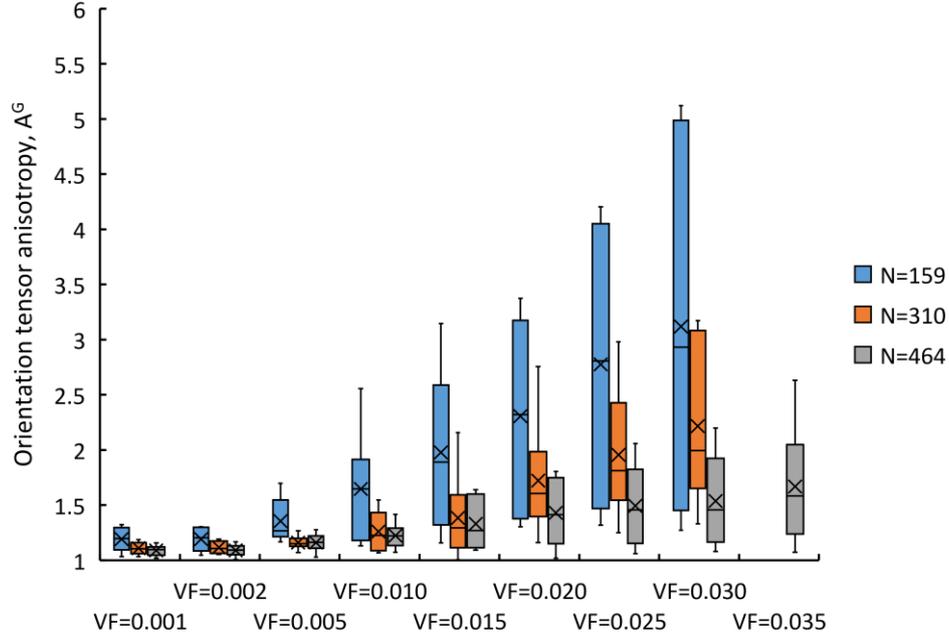

Figure 7 The dependence of the orientation tensor anisotropy for the RVE with 159, 310, and 464 randomly dispersed platelets with the increasing volume fraction of inclusions

On the other side, the degree of anisotropy can be estimated by analyzing the effective anisotropic elastic properties of an RVE as:

$$A^E = \frac{\max(E_x, E_y, E_z)}{\min(E_x, E_y, E_z)}, \tag{9}$$

where $E_x$, $E_y$ and $E_z$ are measures of Young's modulus in three orthogonal directions of an RVE in the local coordinate system. The results for the randomly dispersed platelets with the aspect ratio equal to $10^3$ and the stiffness ratio equal to $10^5$ are presented in the form of a box plot in Figure 8. The absolute values of the elastic anisotropy index $A^E$ cannot be directly compared with the orientation tensor anisotropy $A^G$, however, there is a close correlation between these results. Similar observations were made in Hussein and Kim (2018). The anisotropy is low for dilute compositions when platelets practically do not interact with each other and quickly increases for volume fractions of 0.01 and higher. Comparing this observation with Figures 4–6, it could be noted that for volume fractions below 0.01 most platelets are dispersed randomly and for higher volume fractions compact clusters start to form. As the volume fraction of the filler achieves 0.03, practically all platelets belong to a relatively small number of large clusters, resulting in a highly anisotropic arrangement of platelets. A large number of inclusions is needed to obtain a macroscopically isotropic RVE in this case.

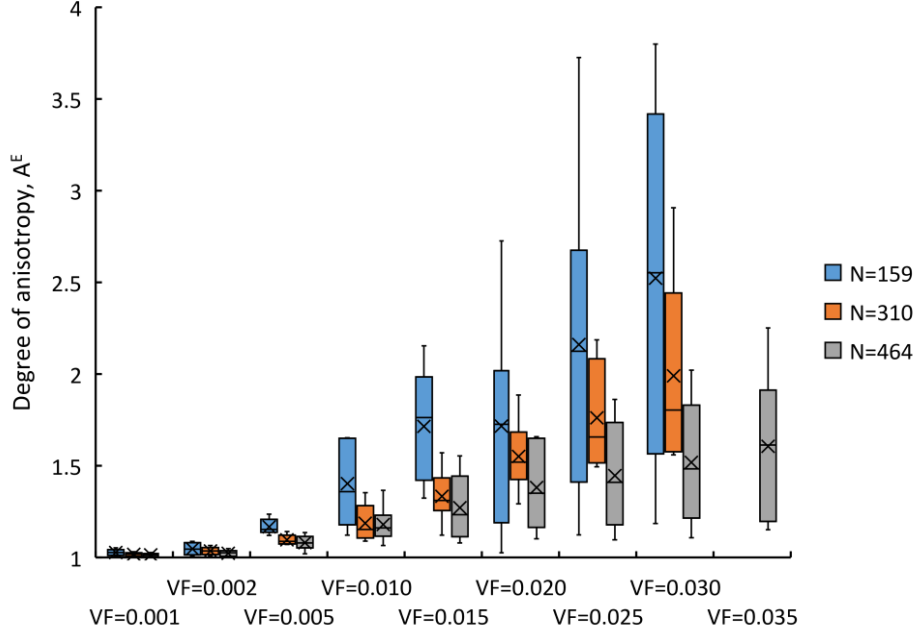

Figure 8 The dependence of the degree of elastic anisotropy for the RVE with 159, 310, and 464 randomly dispersed platelets with the increasing volume fraction of inclusions

## 3.3 Two-step homogenization procedure

Assuming, that each realization of an RVE was obtained by a random process, a macroscopically isotropic elastic behavior could be expected for a sufficiently large RVE. For an RVE with a finite number of particles obtained effective properties are generally anisotropic. A naïve approach of using an average of three stiffness constants (Young's or shear moduli) measured in three orthogonal directions works well for an almost isotropic RVE. However, as was shown above, for a condensed system of platelets grouped into clusters an impractically large number of particles is required to generate a macroscopically isotropic RVE. In the case of a small number of inclusions, a few large clusters of almost parallel platelets define the preferred direction in an RVE, resulting in a severe anisotropy of the effective elastic properties. The results of numerical simulations show, that taking the average of the three Young's or shear moduli measured along the edges of an RVE is not the best strategy, resulting in a large scatter of the effective elastic properties for different realizations of the RVE.

A two-step homogenization procedure was proposed to estimate the isotropic effective properties of an RVE with randomly oriented platelets, which eliminates this problem and produces more accurate and stable results. The idea is to treat the body composed of many generally anisotropic RVE with homogenized properties as a polycrystalline material where each RVE has an arbitrary orientation. Estimation of the effective properties of the polycrystalline material is a classical problem that usually is solved by taking an orientational average of stiffness or compliance matrices of the investigated material. Closed-form expressions for effective shear and bulk moduli of an arbitrary polycrystalline material were derived by Voigt and Reuss (Hill, 1952) and can be written in the form:

$$9K_V = (c_{11} + c_{22} + c_{33}) + 2(c_{12} + c_{23} + c_{31}) \qquad (10)$$

$$15G_V = (c_{11} + c_{22} + c_{33}) - (c_{12} + c_{23} + c_{31}) + 3(c_{44} + c_{55} + c_{66}) \qquad (11)$$

$$1/K_R = (s_{11} + s_{22} + s_{33}) + 2(s_{12} + s_{23} + s_{31}) \tag{12}$$

$$15/G_R = 4(s_{11} + s_{22} + s_{33}) - 4(s_{12} + s_{23} + s_{31}) + 3(s_{44} + s_{55} + s_{66}) \tag{13}$$

where $c_{ij}$ and $s_{ij}$ are components of stiffness and compliance matrices, respectively, $K$ is the bulk modulus, and the subscripts "V" and "R" refer to the Voigt and Reuss approximations. It is interesting to note, that the same result was obtained by searching the isotropic matrix closest to the given anisotropic stiffness or compliance matrices in the Frobenius norm sense (Norris, 2006).

Hill (1952) has proven that the Voigt and Reuss approximations for the effective shear and bulk moduli, $G^V, K^V, G^R$ and $K^R$, are the upper and lower bounds, respectively, and suggested that the true value should lie between them. He proposed to use the arithmetic mean as an approximation of the true effective properties of a polycrystalline material with randomly oriented grains:

$$K_H = {}^1/_2 (K_V + K_R) \tag{14}$$

$$G_H = {}^1/_2 (G_V + G_R) \tag{15}$$

Finally, the two-step homogenization procedure can be described as the following:

1) the numerical model with the periodic RVE is solved six times to determine the compliance matrix, generally anisotropic, of the homogenized material using Eq. (1) (the stiffness matrix is then obtained as an inverse of the compliance matrix);
2) the orientational averaging scheme is employed for the homogenized stiffness and compliance matrices using Eqs. (10)–(13) and the final isotropic properties of the investigated composite material are found using Hill's approximation, Eqs. (14), (15).

Schematically this procedure is illustrated in Figure 9. The proposed procedure is similar to the one used by Vo et al. (2018), where averaging of only a stiffness matrix was used, which may lead to somewhat overestimated values of the effective stiffness. Christensen (1979) used a similar idea for calculating the upper bound of the effective stiffness properties of a platelet-reinforced composite using an orientation averaging of the stiffness matrix of the layered media.

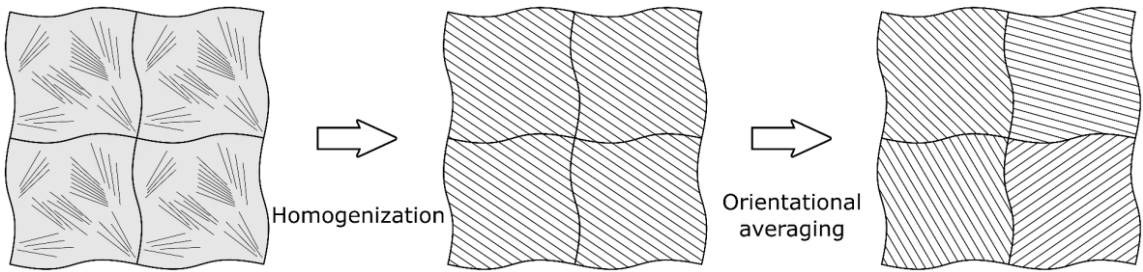

Figure 9 Two-step homogenization procedure: homogenization of a periodic RVE and orientational averaging of a homogenized anisotropic material

Figure 10 shows a mean value and scatter of the effective properties of an RVE with 159, 310, and 464 particles and different volume fractions, calculated using the simple average and the two-step homogenization schemes (six realizations of an RVE were used for each combination of parameters). The platelets with an aspect ratio equal to 1000 and a platelet-to-matrix stiffness ratio equal to $10^5$ were used in these simulations. The results show that even though the mean values within each group with an equal volume fraction of the filler are quite close, the scatter for the simple average method is significantly larger, than by using the two-step homogenization procedure. The scatter in the effective properties of an RVE with 159 particles using the two-step homogenization (yellow bar) is about the same or smaller than the scatter in properties of an RVE with 464 particles using the simple average (gray bar), therefore an RVE with a much smaller number of particles can be used for the same degree of accuracy by applying the two-step homogenization.

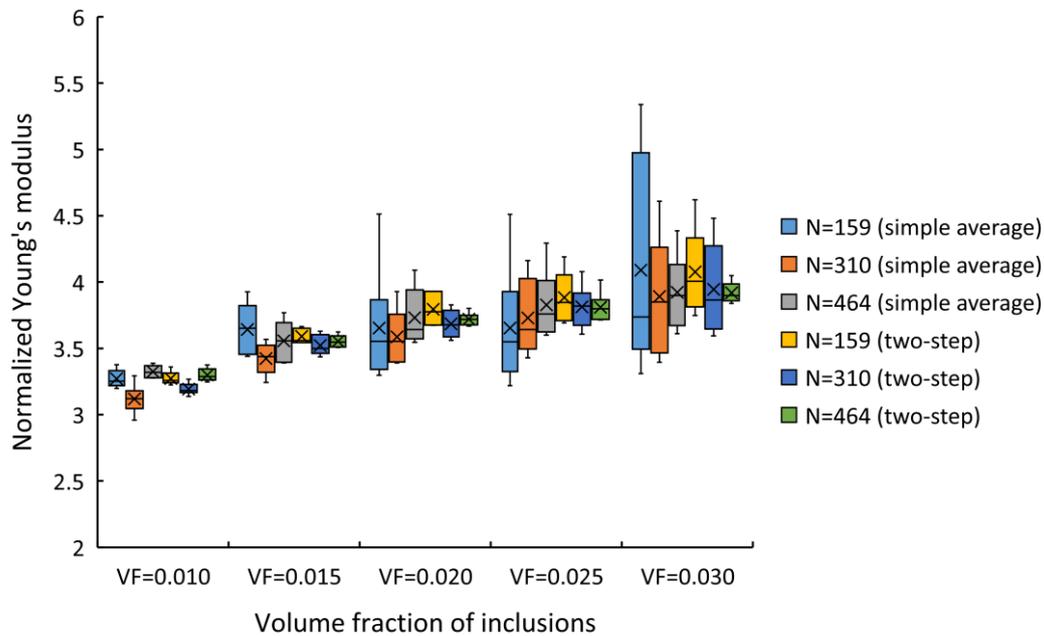

Figure 10 The mean value (×) and the scatter (min, max, and quartiles) of the effective stiffness of the RVE with a different number of particles and different volume fractions calculated using simple average and two-step homogenization schemes

### 3.4 Effect of formation of clusters on the effective properties of composite

The effect of a clusterization on the effective elastic properties of a composite material reinforced by thin stiff platelets was estimated numerically using an RVE with 464 randomly dispersed platelets with an aspect ratio equal to 1000 and a platelet-to-matrix stiffness ratio ranging from 30 to $10^5$.

Figure 11 shows a dependence of the effective Young's modulus of the composite material on the volume fraction and normalized stiffness of the filler, $E_f/E_m$. Average values of six different realizations of an RVE (using the two-step homogenization method for each realization) are shown in Figure 11 with error bars marking the standard deviation of the results, however, for most numerical points, the error bars are smaller than the size of the point. The dashed lines present the predictions by the Halpin-Tsai equations. The predictions by the Mori-Tanaka method for these parameters of the RVE were slightly higher (by less than 4%) than the Halpin-Tsai predictions.

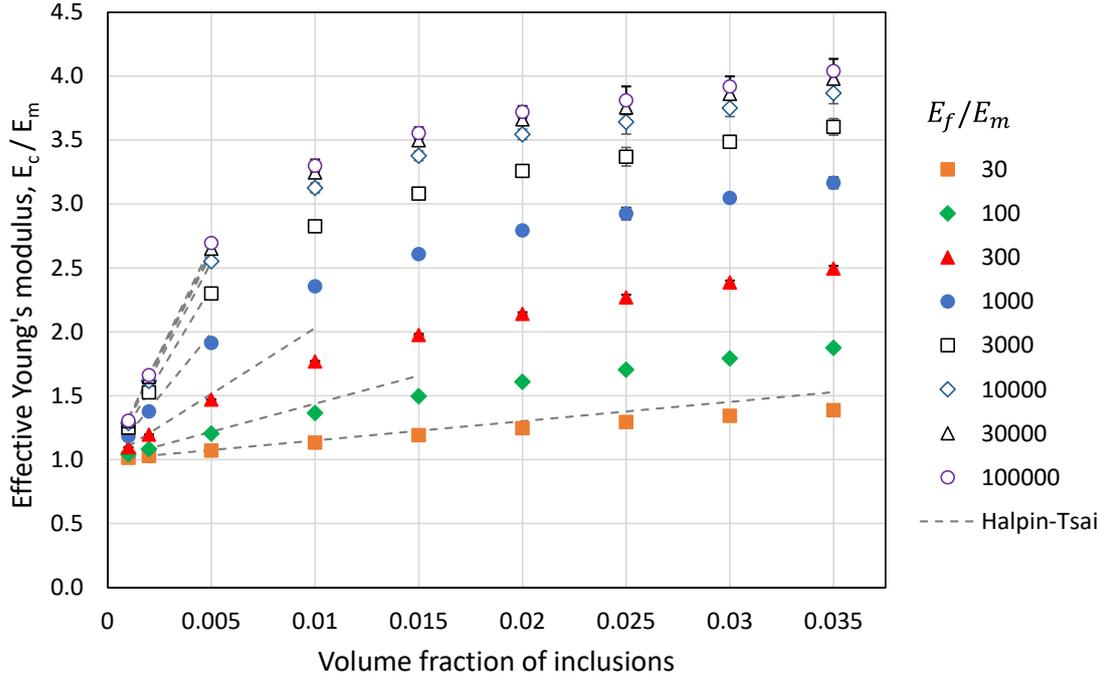

Figure 11 Effective stiffness of the composite material with different volume fractions of reinforcement. The number of platelets in the RVE – 464, aspect ratio of platelets – 1000. Dashed lines – Halpin-Tsai predictions

The results show that the effective stiffness of the composite increases with increasing the stiffness of the filler, however at some point further increase of the platelet's modulus practically does not influence the effective properties of a composite (Sheng et al., 2004). In current simulations with the filler having the aspect ratio equal to 1000, a plateau is achieved at the normalized stiffness of the filler, $E_f/E_m$, equal to $10^4$. At this point, the particle can be treated as an absolutely hard inclusion and further increase of the particle's stiffness practically does not change a strain field inside the matrix.

The influence of a volume fraction of inclusions shows an almost linear trend for low fractions of inclusions, however, a non-linear behavior is observed as the content of the filler increases. The predictions by the Halpin-Tsai equations show a very good agreement with numerical simulations at low fractions of filler, up to 0.5%. The images of the distributions of platelets within an RVE (Figure 6) show that up to the volume fraction of 0.5% the arrangement of platelets is almost perfectly random, which is also supported by the low values of the orientation tensor anisotropy (Figure 7). At the volume fraction of 1%, the first clusters of tightly packed platelets appear, resulting in reducing the reinforcing capacity of inclusions. Similar observations were made in previous studies (Ji et al., 2010). At the same time, the anisotropy of an RVE quickly increases with increasing the volume fraction of a filler, as shown in Figure 8.

## 3.5  Scaling relations for the effective elastic properties

For thin platelets, with an aspect ratio greater than 200-300, simple scaling relations can be used to generalize obtained numerical results for a composite material with an arbitrary stiffness and aspect ratio of reinforcement. Thin platelets are effectively two-dimensional inclusions, for which out-of-plane dimensions and elastic properties can be neglected in many cases. The bending of platelets in the current analyses is quite low and also can be neglected. The major reinforcing factor is the membrane stiffness of inclusions and, therefore, it is possible to replace analyzed platelets with equivalent particles with the same membrane stiffness, and the effective properties of new material should not change significantly. Based on this assumption, it is sufficient to build a numerical model and analyze the system with only one particular aspect ratio of inclusions. Then for an arbitrary material reinforced by platelets with the aspect ratio $AR$, volume fraction $V_f$, and stiffness $E_f$, it is necessary to replace the platelets with equivalent particles with normalized volume fraction $\hat{V}_f$ and stiffness $\hat{E}_f$, using relations (16)-(18), to use numerical results from Figure 11, obtained for inclusions with base aspect ratio $AR_0 = 1000$.

Scaling factor: $k = AR_0/AR$ (16)

Normalized volume fraction: $\hat{V}_f = V_f/k$ (17)

Normalized Young's modulus: $\hat{E}_f = kE_f$ (18)

These relations allow minimizing the number of expensive simulations during preliminary parametric studies.

## 4 Conclusions

The effect of the formation of compact dense clusters on the effective elastic properties of the platelet-reinforced composites was studied using a periodic RVE with randomly dispersed particles. A hybrid 2D/3D finite element model, where thin inclusions are modeled using 2D shell elements, sharing nodes and faces with 3D solid elements, representing the surrounding matrix, was used to reduce the computational time.

The optimal size of the RVE was estimated using three generations of RVE with 159, 310, and 464 inclusions. The results show that a significant scatter in the calculated effective properties was observed at high volume fractions of the filler when most inclusions belong to a relatively small number of compact clusters. To eliminate this problem, the two-step homogenization scheme was proposed, which allows obtaining stable results with a smaller number of particles.

The effect of compact clusters of nearly parallel platelets on the effective elastic properties of the composite material was studied using an RVE with 464 randomly dispersed inclusions with the aspect ratio equal to 1000 for different volume fractions and inclusion-to-matrix stiffness contrasts. A good agreement was observed between analytical predictions using the Halpin-Tsai equations and numerical results for dilute systems, however, at higher volume fractions of the filler, as the clusters of nearly parallel platelets start to form, a numerical solution should be used. Simple scaling relations were proposed to generalize obtained numerical results for platelets with different aspect ratios.

## Acknowledgments

This work was developed under the M-era.Net research project titled NANO2COM – Advanced Polymer Composites Filled with Novel 2D Nanoparticles. The research was funded by grant No. 1.1.1.5/ERANET/18/02 from the Latvian State Education Development Agency.

# References

Advani, S.G., Tucker III, C.L., 1987. The use of tensors to describe and predict fiber orientation in short fiber composites. J. Rheol. 31, 751-784.

Affdl, J.C.H., Kardos, J.L., 1976. The Halpin-Tsai equations: A review. Polymer Engineering & Science 16, 344-352.

Benveniste, Y., 1987. A new approach to the application of Mori-Tanaka's theory in composite materials. Mech. Mater. 6, 147-157.

Bock, H., Blümling, P., Konietzky, H., 2006. Study of the micro-mechanical behaviour of the Opalinus Clay: an example of co-operation across the ground engineering disciplines. Bulletin of Engineering Geology and the Environment 65, 195-207.

Christensen, R.M., 1979. Isotropic Properties of Platelet-Reinforced Media. J. Eng. Mater. Technol. 101, 299-303.

Coelho, D., Thovert, J.-F., Adler, P.M., 1997. Geometrical and transport properties of random packings of spheres and aspherical particles. Physical Review E 55, 1959.

Delhorme, M., Jönsson, B., Labbez, C., 2012. Monte Carlo simulations of a clay inspired model suspension: the role of rim charge. Soft Matter 8, 9691-9704.

Ebrahimi, D., Whittle, A.J., Pellenq, R.J.-M., 2016. Effect of polydispersity of clay platelets on the aggregation and mechanical properties of clay at the mesoscale. Clays Clay Miner. 64, 425-437.

Feder, J., 1980. Random sequential adsorption. Journal of theoretical biology 87, 237-254.

Figiel, Ł., Buckley, C.P., 2009. Elastic constants for an intercalated layered-silicate/polymer nanocomposite using the effective particle concept – A parametric study using numerical and analytical continuum approaches. Computational Materials Science 44, 1332-1343.

Geuzaine, C., Remacle, J.-F., 2009. Gmsh: A 3-D finite element mesh generator with built-in pre- and post-processing facilities. International Journal for Numerical Methods in Engineering 79, 1309-1331.

Giannelis, E.P., 1996. Polymer layered silicate nanocomposites. Adv. Mater. 8, 29-35.

Gusev, A.A., 1997. Representative volume element size for elastic composites: a numerical study. Journal of the Mechanics and Physics of Solids 45, 1449-1459.

Halim, J., Kota, S., Lukatskaya, M.R., Naguib, M., Zhao, M.-Q., Moon, E.J., Pitock, J., Nanda, J., May, S.J., Gogotsi, Y., Barsoum, M.W., 2016. Synthesis and Characterization of 2D Molybdenum Carbide (MXene). Adv. Funct. Mater. 26, 3118-3127.

Hbaieb, K., Wang, Q., Chia, Y., Cotterell, B., 2007. Modelling stiffness of polymer/clay nanocomposites. Polymer 48, 901-909.

Hill, R., 1952. The elastic behaviour of a crystalline aggregate. Proceedings of the Physical Society. Section A 65, 349.

Hussein, A., Kim, B., 2018. Graphene/polymer nanocomposites: the active role of the matrix in stiffening mechanics. Compos. Struct. 202, 170-181.

Ji, X.-Y., Cao, Y.-P., Feng, X.-Q., 2010. Micromechanics prediction of the effective elastic moduli of graphene sheet-reinforced polymer nanocomposites. Modell. Simul. Mater. Sci. Eng. 18, 045005.

Kilikevičius, S., Kvietkaitė, S., Žukienė, K., Omastová, M., Aniskevich, A., Zeleniakienė, D., 2020. Numerical investigation of the mechanical properties of a novel hybrid polymer composite reinforced with graphene and MXene nanosheets. Computational Materials Science 174, 109497.

Lan, T., Kaviratna, P.D., Pinnavaia, T.J., 1995. Mechanism of clay tactoid exfoliation in epoxy-clay nanocomposites. Chem. Mater. 7, 2144-2150.

Li, S., 2001. General unit cells for micromechanical analyses of unidirectional composites. Composites Part A: applied science and manufacturing 32, 815-826.

Li, S., Wongsto, A., 2004. Unit cells for micromechanical analyses of particle-reinforced composites. Mech. Mater. 36, 543-572.

Messersmith, P.B., Giannelis, E.P., 1994. Synthesis and characterization of layered silicate-epoxy nanocomposites. Chem. Mater. 6, 1719-1725.

Monastyreckis, G., Mishnaevsky, L., Hatter, C.B., Aniskevich, A., Gogotsi, Y., Zeleniakiene, D., 2020. Micromechanical modeling of MXene-polymer composites. Carbon 162, 402-409.

Mori, T., Tanaka, K., 1973. Average stress in matrix and average elastic energy of materials with misfitting inclusions. Acta Metall. 21, 571-574.

Mortazavi, B., Bardon, J., Ahzi, S., 2013a. Interphase effect on the elastic and thermal conductivity response of polymer nanocomposite materials: 3D finite element study. Computational Materials Science 69, 100-106.

Mortazavi, B., Hassouna, F., Laachachi, A., Rajabpour, A., Ahzi, S., Chapron, D., Toniazzo, V., Ruch, D., 2013b. Experimental and multiscale modeling of thermal conductivity and elastic properties of PLA/expanded graphite polymer nanocomposites. Thermochim. Acta 552, 106-113.

Naguib, M., Mochalin, V.N., Barsoum, M.W., Gogotsi, Y., 2014. 25th Anniversary Article: MXenes: A New Family of Two-Dimensional Materials. Adv. Mater. 26, 992-1005.

Norris, A., 2006. The isotropic material closest to a given anisotropic material. Journal of Mechanics of Materials and Structures 1, 223-238.

Peng, R., Zhou, H., Wang, H., Mishnaevsky Jr, L., 2012. Modeling of nano-reinforced polymer composites: Microstructure effect on Young's modulus. Computational Materials Science 60, 19-31.

Segurado, J., Llorca, J., 2002. A numerical approximation to the elastic properties of sphere-reinforced composites. Journal of the Mechanics and Physics of Solids 50, 2107-2121.

Sheng, N., Boyce, M.C., Parks, D.M., Rutledge, G., Abes, J., Cohen, R., 2004. Multiscale micromechanical modeling of polymer/clay nanocomposites and the effective clay particle. Polymer 45, 487-506.

Spanos, K.N., Georgantzinos, S.K., Anifantis, N.K., 2015. Mechanical properties of graphene nanocomposites: A multiscale finite element prediction. Compos. Struct. 132, 536-544.

Van Es, M., 2001. Polymer-clay nanocomposites. Delft.

Vila-Chã, J.L.P., Ferreira, B.P., Pires, F.M.A., 2021. An adaptive multi-temperature isokinetic method for the RVE generation of particle reinforced heterogeneous materials, Part I: Theoretical formulation and computational framework. Mech. Mater. 163, 104069.

Vo, V.S., Nguyen, V.-H., Mahouche-Chergui, S., Carbonnier, B., Naili, S., 2018. Estimation of effective elastic properties of polymer/clay nanocomposites: A parametric study. Composites Part B: Engineering 152, 139-150.

Wang, K., Chen, L., Wu, J., Toh, M.L., He, C., Yee, A.F., 2005. Epoxy nanocomposites with highly exfoliated clay: mechanical properties and fracture mechanisms. Macromolecules 38, 788-800.

Widom, B., 1966. Random sequential addition of hard spheres to a volume. The Journal of Chemical Physics 44, 3888-3894.

Zilg, C., Mülhaupt, R., Finter, J., 1999. Morphology and toughness/stiffness balance of nanocomposites based upon anhydride-cured epoxy resins and layered silicates. Macromol. Chem. Phys. 200, 661-670.